\documentclass[letterpaper,twocolumn,10pt]{article}
\usepackage{usenix-2020-09}% 
\usepackage{tabularx}
\usepackage{cleveref}
\usepackage{xspace}
\usepackage{graphicx}
\usepackage{xcolor}

\begin{document}

\newcommand{\snp}{S\&P\xspace}
\newcommand{\snpns}{S\&P}
\title{Secure and Private Spatial Sharing for Mixed Reality Remote Collaboration in Enterprise Settings}

%%%%%%%%%%%%%%%% Authors' Info %%%%%%%%%%%%%%%%%
\author{\rmfamily\mdseries\upshape
  Mengyu Chen\textsuperscript{1}\textsuperscript{*} \quad
  Youngwook Do\textsuperscript{1}\textsuperscript{*} \quad
  Feiyu Lu\textsuperscript{1}\textsuperscript{*} \quad
  Kaiming Cheng\textsuperscript{2} \quad
  Blair MacIntyre\textsuperscript{1}\\[0.5em]
  \textsuperscript{1}JPMorganChase, New York, NY, USA \quad
  \textsuperscript{2}University of Washington, Seattle, WA, USA\\[0.4em]
  % \email{\{mengyu.chen, youngwook.do, feiyu.lu, blair.macintyre\}@jpmchase.com} \\
  % \email{kaimingc10@gmail.com}\\
  \textsuperscript{*}Equal contribution
}

% \begin{IEEEkeywords}
% Access Control, Data Privacy, Enterprise Security, Internet of Things, Mixed Reality, Remote Collaboration, Spatial Sharing
% \end{IEEEkeywords}


\maketitle
\pagestyle{empty} 

\begin{figure*}[h!]
    \centering
    \includegraphics[width=\linewidth]{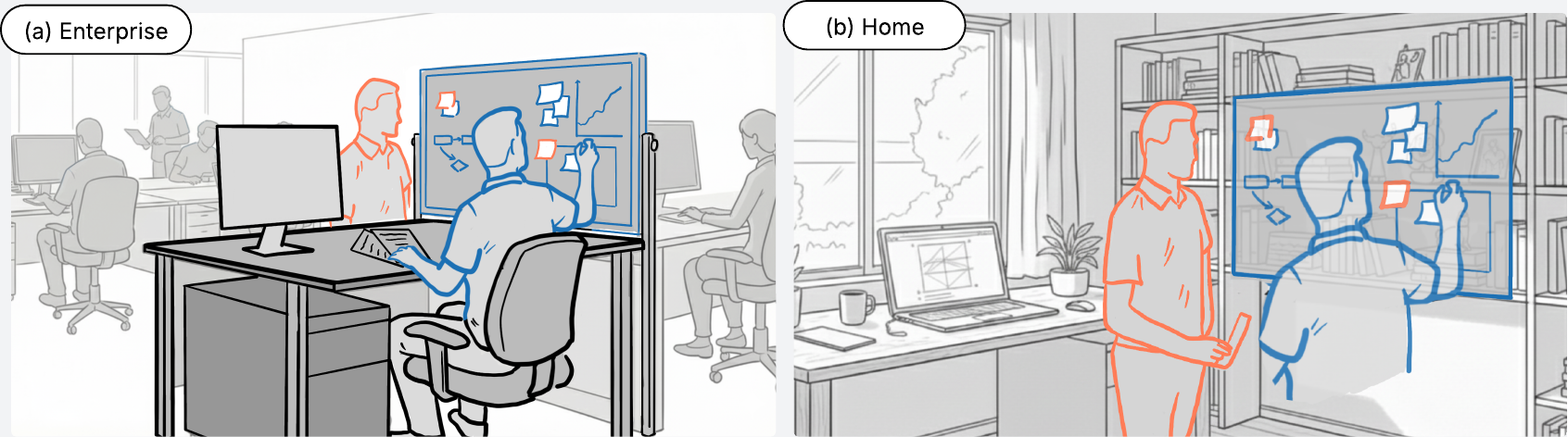}
    \caption{A demonstration of MR spatial sharing between enterprise and home settings. (a) An employee Bob (blue) who works in an enterprise office shares a physical whiteboard with Carol (red) who is a client at home. Sensitive information such as monitors, documents on the table are opted out from sharing due to company policy. (b) Carol sees the whiteboard and shares his sticky notes with Bob. Private environmental information at his home are opted out from sharing with Bob by Carol.}
    \label{fig:teaser_image}
\end{figure*}

\section*{Abstract} 
Mixed Reality (MR) technologies are increasingly adopted by enterprises to enhance remote collaboration, enabling users to share real-time views of their physical environments through head-mounted displays (HMDs). While MR spatial sharing offers significant benefits, it introduces complex security and privacy risks, particularly in balancing employee collaboration needs with enterprise data protection requirements across office and personal spaces. This paper investigates these challenges through formative interviews with employees and expert consultations with professionals in cybersecurity, IoT, technology risk, and corporate legal domains. We present a conceptual framework for secure MR spatial sharing in enterprise contexts and identify critical concerns and requirements for system design. Based on our findings, we offer actionable recommendations to guide the development of secure and privacy-preserving MR spatial sharing solutions for future enterprise deployments.

\par\vspace{0.75\baselineskip}
\noindent\textbf{Keywords—} Access Control, Data Privacy, Enterprise Security, Internet of Things,\\
Mixed Reality, Remote Collaboration, Spatial Sharing

\section{Introduction}

% 1. MR collaboration got a lot of interest from enterprise. Physical spatial sharing is also introduced as a novel feature. 
% 2. there is snp issues due to camera capture
%%% 2-1. why is it different in enterprise settings compared to others
% 3. to understand a broad picture of snp, we did expert interviews. 
%%% 3-1. we used Probe as an inspiration to spark discussion, and found results. 
% 4. contributions + conclusion

% 1. trend of view sharing
Enterprises are increasingly interested in exploring the use of Mixed Reality (MR) technology to improve remote collaboration. Many commercially available Head-Mounted Displays (HMDs) allow wearers to share video and space captured from their point-of-view during collaborative experiences. This feature allows collaborators to see the user's activities or physical surroundings regardless of whether they are in a different office or working from home. For example, during a technical support call, a user might use their HMD to share the view of their physical surrounding including the space and objects for a technician to provide instructions from a first-person view \cite{bun2021using, cheng2024spatialprivacy}. During a physical training call, a remote expert may share their perspectives to facilitate learning.

% 2. sp risk about view sharing, two dimensions risk - employee versus enterprise, actual physical space vs personal space, 2x2 situation leading to different requirements 
Despite the potential benefits of view-sharing of an HMD wearer's physical surroundings, this feature can create security and privacy (\snpns) risks in enterprise settings \cite{rajaram2023eliciting}. Specifically, the risks stem from potential conflicts in two dimensions: \textbf{the functional needs of different entities} (employees' need to collaborate efficiently versus enterprise need to prevent data leakage \cite{beautement2016} and \textbf{physical space context} (office space versus personal space). 
% As Naeini et al. suggest, each setup requires different \snp considerations.  
% Additionally, remote collaboration setups have made it possible to work in different physical locations (e.g., company office, personal space), which may need the different sets of considerations to reduce \snp risks. 
Addressing these usage risks requires tailoring \snp requirements to the specific needs of different stakeholders and the distinct attributes of various physical spaces.
For example, in an office setting, enterprise policy may be set to restrict streaming the camera view of any physical space or object to reduce the risk of a data breach even if an employee tries to share their physical office whiteboard via the HMD camera to collaborators in a remote meeting. Similarly, an employee working from home in their personal space may need to share a view of ad hoc physical objects, such as hardware and instruments, with attendees in the office.
% business could be at \snp risk when an employee unintentionally record sensitive space or objects placed in a physical office and stream them during a call to clients who are unauthorized to access any internal data. 
However, recording and streaming a view of an employees' personal space or items in that space may violate their personal \snp even though the recording would not directly harm enterprise \snp. 

In this paper, we aim to explore the spectrum of challenges to designing \snp-conscious MR spatial sharing systems by understanding complex considerations for different contexts associated with the two dimensions mentioned above. We have identified two challenges doing so. First, \snp risks are highly context-dependent \cite{nissenbaum2009privacy}, varying with the actors involved, the physical environment, and the goals of the activity, yet prior work has examined these dynamics primarily in desktop and mobile ecosystems, not in MR settings where enterprise and employee contexts intersect \cite{guzman2019, sajid2025}. Second, MR adoption in enterprise is at the early stage\cite{jalo2022extended, fortune2025xr}, resulting in a lack of understanding of the context where employees want to use MR spatial sharing and evidence regarding \snp risks associated with enterprise use cases \cite{varasainen2021}.

These challenges motivated us to think about \snp risks in the enterprise settings. As a first step, we aimed to understand which contexts employees envision for MR spatial sharing. Given MR adoption is still at an early stage, we identified employee's spatial sharing use cases on commonly-used video calls, and used the associated \snp concerns as reference. Second, we wanted to understand which \snp features are expected in MR spatial sharing. To do so, we ran a formative interview study with enterprise employees (N=10). In summary, employees used spatial sharing as an ad hoc, simple way to show items nearby and brainstorm ideas (e.g, whiteboard). Also, while they envisioned the similar use cases in MR, they expressed the need to control content filtering and video access depending on whether the physical space was personal or in the office. This finding highlighted the importance of flexible control mechanisms for spatial sharing, tailored to the context of use and the type of physical space involved. 

% \todo{one more paragraph, this domain has not been well studies, it is important to understand the scope of the problem, provide suggestions about this problem domain. We can mention, there wouldn't be one-size-fit-all solution especially because of the enterprise space, so it is hard to generalize a problem without studying carefully about it....}

% 3. what we do, how we use probe / reason for using probe.

% However, there has been little studied regarding the holistic view of \snp risks for remote collaboration settings in enterprise. 
Once we had an initial understanding of the contexts where employees want to use MR spatial sharing, we aimed to understand both employee's and enterprise's perspectives based on those contexts. 
To do so, we ran an interview study with domain experts in industry to understand \snp risks of MR spatial sharing in the enterprise context and how to address them. For the interview, we designed a conceptual framework as a probe that illustrates data flow of a spatial sharing system, aiming to  mitigate the risk of disclosing private and confidential information via a HMD device's cameras. We used this probe during our interview as a study method, inspired by Gaver et al.\cite{gaver1999design}, to initiate and amplify the discussion regarding \snp risks. This way, participants were able to have a clear understanding of the system structure and ask questions based on the structure of the framework before the system implementation. Based on the discussions with the participants, we identified key factors to consider when developing an MR spatial sharing system. In summary, while participants mentioned the need to make the system usable for employees, they emphasized the need to optimize the balance toward the enterprise needs to reduce \snp risks and financial burden to deploy and manage the system. In short, our contributions include:

\begin{itemize} 
    \item We present the design of a probe framework for MR spatial sharing tailored to enterprise environments, enabling secure and policy-driven multi-user collaboration. 
    \item We conduct an expert interview study with seven professionals specializing in cybersecurity, IoT research, technology risk control, and corporate legal domains. Through these interviews, we identify critical \snp concerns, challenges, and feature requirements for secure and usable MR spatial sharing in enterprise settings. 
    \item Based on our findings, we propose six actionable design recommendations to guide practitioners in implementing secure and privacy-preserving MR spatial sharing for future enterprise deployments. 
    
\end{itemize}

\section{Background and Motivation}

We begin with background in spatial sharing technologies, MR remote collaboration methods, and prior work in MR environments' \snp to contextualize our motivation of this work.

\subsection{\snp Risks of Video Conferencing}

% 1. 
Video conferencing has enabled co-workers to have face-to-face meetings even if they are not physically co-located, allowing them to communicate with each other through their webcams from different locations \cite{mantei1991experiences, gaver1992affordances}. However, this setup poses \snp risks to users. For example, a webcam could be covertly accessed with software manipulation and used to monitor a user without their knowledge or consent \cite{brocker2014iseeyou}. In order to prevent such unauthorized access, it is recommended to install the latest security update for a user's software \cite{ion2015no}. 

Despite software updates, there remains the \snp risks to users due to human error. Specifically, users may not notice sensors are activated and capturing sensitive information. For example, nowadays video conference tools feature mute and unmute options for a microphone and a camera. However, users might unintentionally unmute their camera or microphone in video conference, allowing other attendees to see or hear their private activities \cite{prange2022saw}. To mitigate such risks, webcams are designed to provide visuals that indicate mute/unmute status of sensors (e.g., when a camera is activated, a green indicator is displayed near it or on the screen) \cite{portnoff2015somebody}. Additionally, users have started covering their webcam with a physical cover (e.g., tape) to physically block the webcam view even if the webcam access is compromised \cite{machuletz2018webcam} although they often forget to cover a webcam as the webcam covering relies on human memory \cite{do2021smart}. 

On the other hand, users may be aware of a sensors' activation yet may not notice it could accidentally capture more data than users' intent. For instance, while a user is having a video call in their open-desk office with an external client who is unauthorized to learn certain kinds of internal information, other employees' monitors could be visible on the users' background allowing the client to see them. Video conference providers have started deploying virtual background features that only show a user by occluding their background. 

% In this paper, we focus on an \snp risks of HMD device's external cameras. Specifically, a unique factor of the HMD device compared to a webcam-based video conference is its spatial awareness. The HMD device uses spatial awareness to display virtual content overlays on top of their passthrough view, which often requires the recording by a series of external cameras. As the HMD is using cameras yet different setups with the aforementioned video conference setups, we aim to identify the \snp risk by those cameras and the design considerations to build privacy-aware applications with the cameras.
In this paper, we focus on the \snp risks associated with an HMD's external cameras. Unlike fixed webcams used in video conferencing, HMDs are worn and mobile; to maintain spatial awareness, they continuously sense and map the environment as the wearer moves, rendering virtual overlays in the passthrough view. This capability typically relies on one or more outward-facing cameras that capture the physical scene throughout a session. Given these differences from conventional webcams, we aim to identify the \snp risks posed by these cameras and outline design considerations for privacy-aware application development.

\subsection{\snp Practice for Smart Devices in Workplace}

% -Recourse: sensing only what you need. ---  Privacy by Design (Marc Langheinrich)

% 1. Enterprises are worried about S&P. So, they enforce policies for device usage and the devices are regulated in compliance with the devices. 
% 2. As smart office emerges, the spectrum of the policy expants to cover smart devices that have sensors like cameras. 
% 3. however, this policy may not need to be applied for some cases because it depends on contexts. Any current examples. 
% 4. also if it's difficult to use, people find a way to detour, leading to a bad consequence. 
% 5. We are doing something similar for a new device, XR HMD. 

Data \snp is crucial for enterprises as their data could include the companies' sensitive information such as employee data. Despite the importance of data \snp, data breaches could occur for various reasons including security system failures or human error in technology usage. Even if fixing the security system failures could secure the system, it would still not be effective if a user makes bad decisions while using the system \cite{kraemer2007human}.
Owing to that, enterprises often apply policy enforcement for their technology assets to protect the company's \snp. Moreover, the International Organization for Standardization (ISO) has published ISO/IEC 27002 that lists various guidelines for companies to apply to improve their \snp practice \cite{iso27002_2022}. 

As ``smart'' devices that have sensors embedded and wireless networks have become part of work assets, policy enforcement has covered data collection by sensors and online data transfers. For example, companies often enforce virtual private network (VPN) connection for employees to authenticate and join internal video conference rooms. Otherwise, the video conference access would be restricted without the VPN connection. This way, collected sensor data such as webcam audio/video can be securely streamed for a video conference, protected by security layers including the corporate VPN. 

Besides the enforceable policy regarding technology, like restricting devices' features, companies expect their employees to follow ``good'' practices shaped to protect confidential information from being disclosed publicly. Additionally, \snp practices often are applied differently depending on contexts. For instance, working remotely with a laptop from public spaces may not be considered good practice since someone could shoulder-surf and see confidential information about the user's organization, which is of no concern if working from a private space such as home. Therefore, \snp compliance can be achieved by both the policy enforcement and the context-dependent good practice. 

In this work, we explore various design parameters that encompass policy enforcement and good practices for enterprises in the adoption of new technologies, with a focus on HMD devices. We, further, delve into how the design parameters could be utilized for different contexts where employees are situated.

\subsection{Remote Collaboration in Mixed Reality}

%Signify why remote collab and why spatial sharing is common
The primary objectives of real-time remote collaboration systems are to foster a sense of togetherness and to reduce communication barriers. MR introduces innovative paradigms to accomplish these goals by providing a comprehensive understanding of users' spatial pose with respect to shared content and their surrounding environment. Over the past few decades, researchers from the fields of HCI, CSCW, and MR have explored the potential of HMDs to enhance remote assistance, training, and collaboration ~\cite{fidalgo2023survey,billinghurst1999collaborative,gurevich2015design, feick2018, piumsomboon2019effects}. Existing literature frequently refers to sharing one's physical workspace with a remote party as ``view-sharing''~\cite{lee20,kim2018effect,lee2017mixed}, which has drawn significant interest from the community over the past decades.  Early work by Adcock et al. developed a remote guidance system that uses depth fusion of a workers space to remote experts \cite{adcock2013remotefusion}. Lee et al. developed a MR collaboration system in which a trainer who remotely assisted a trainee's task by seeing a real-time panorama capture of the trainee's physical space in an HMD \cite{lee17}. Teo et al. shared 360 panorama capture and 3D reconstruction of the local user's with remote MR helpers to facilitate communication and guidance \cite{teo19}. This body of work demonstrates the significant potential of HMDs to support remote collaboration in professional and enterprise settings by capturing and sharing users' surrounding environments. Recent industry offerings have made spatial capturing and sharing low-friction and widely-available to everyday consumers. Notably, Meta’s Hyperscape Capture, released in November 2025, enables users to generate 3D Gaussian Splatting representations of their environments with a Quest device and share them with others~\cite{meta-hyperscape}.

With the continuous advancement and proliferation of the MR device market, we foresee an increasing integration of HMDs into everyday enterprise workflows. As these devices become more accessible and capable, organizations are likely to leverage MR for a broader range of collaborative and operational tasks \cite{varasainen2021, irlitti2024}. However, this growing adoption also brings new challenges, particularly in terms of security, privacy, and policy compliance. Recent studies have begun to highlight concerns around the capture and sharing of sensitive information, both intentionally and unintentionally, through MR devices in physical workspaces ~\cite{sajid2025, roesner2021security, lebeck2018towards, ruth2019secure}. Additionally, Do et al. studied methods to communicate the HMD's passthrough recording to bystanders co-located in a physical space \cite{do2023vice}. Rajaram et al. proposed the framework designed to help balance between user experiences on HMD usages and privacy needs of the co-located people including bystanders in a physical space \cite{rajaram2025privacy}.

However, these concerns have not been well-articulated in enterprise settings, where company-wide policies and employees' needs surface new regulations and conflicts. These factors remain uncovered by previous work. In this research, we attempt to fill this gap by conducting formative studies with corporate workers and evaluating a framework with experts from engineering, compliance, and risk backgrounds.

\subsection{Spatial Sharing Technologies}

% The spatial data is Mapping is the process of creating a 3D model, usually in the form of 3D point clouds - 

% Simultaneous Localization and Mapping (SLAM)
% Structure-from-Motion (SfM) -

% Gaussian Splatting, Nerf,

% Maybe talk about some work on privacy for 2D remote collaboration - 

% These constitute the foundation for leveraing mixed reality for remote meeting and collaboration.

Early work about space capturing and sharing typically involve additional hardware such as depth and 360-degree cameras \cite{teo19,adcock2013remotefusion}. The developments in depth estimation \cite{huang2024reversion} and Simultaneous Localization and Mapping (SLAM) accelerated the process of capturing 3D spaces, though these methods require a considerable amount of time to fully capture and process the 3D scanning of a scene without accurate representations of mesh structures and textures. Until more recently, with the advancements in neural network and radiance fields, approaches such as the Neural Radiance Fields (NeRF) \cite{mildenhall2021nerf} and 3D Gaussian Splatting \cite{kerbl20233d} further accelerated the 3D reconstruction process. 3D space capturing could be created from single or multi-view of 2D images, rendered in photorealistic point clouds, which can then be complemented or edited in real time with generative AI \cite{hong2024lrmlargereconstructionmodel, chenyiwen2024}. These technologies have considerably reduced the cost of capturing and sharing target environments. Notably, some device manufacturers such as the Apple's Vision Pro, includes support for capturing and sharing 3D photos, as well as for applications to scan, track, and share real-world objects and environments \footnote{https://developer.apple.com/documentation/visionos/implementing-object-tracking-in-your-visionos-app}. These developments open up new possibilities for remote assistance, content sharing, and collaboration in enterprise workplace settings with MR technologies.

% Mention the elevated risks of MR spatial sharing comapred to 2D mediums such as zoom. Focus more on enterprise level
However, spatially capturing and sharing the user's surrounding  in MR comes with inherent \snp concerns and risks that existing \snp protocols may fall short \cite{rajaram2023eliciting, kablo2025}. As compared to traditional video conferencing tools, we identified the following unique challenges in MR: 1) cameras for video conferencing could be turned off or physically blocked to opt out from sharing any content, while the egocentric cameras of MR devices have to run continuously in order to localize the users and register virtual content persistently \cite{gallardo2023speculative}, 2) as compared to video conference calls which happen mostly in stationary settings, MR collaboration sessions are often situated in the 3D physical environment, which involves wide range of motion and dynamic user activities \cite{warin2025}. As such, user activities become vulnerable to potential \snp breaches, both intentionally and unintentionally.

\begin{figure*}
    \centering
    \includegraphics[width=\textwidth]{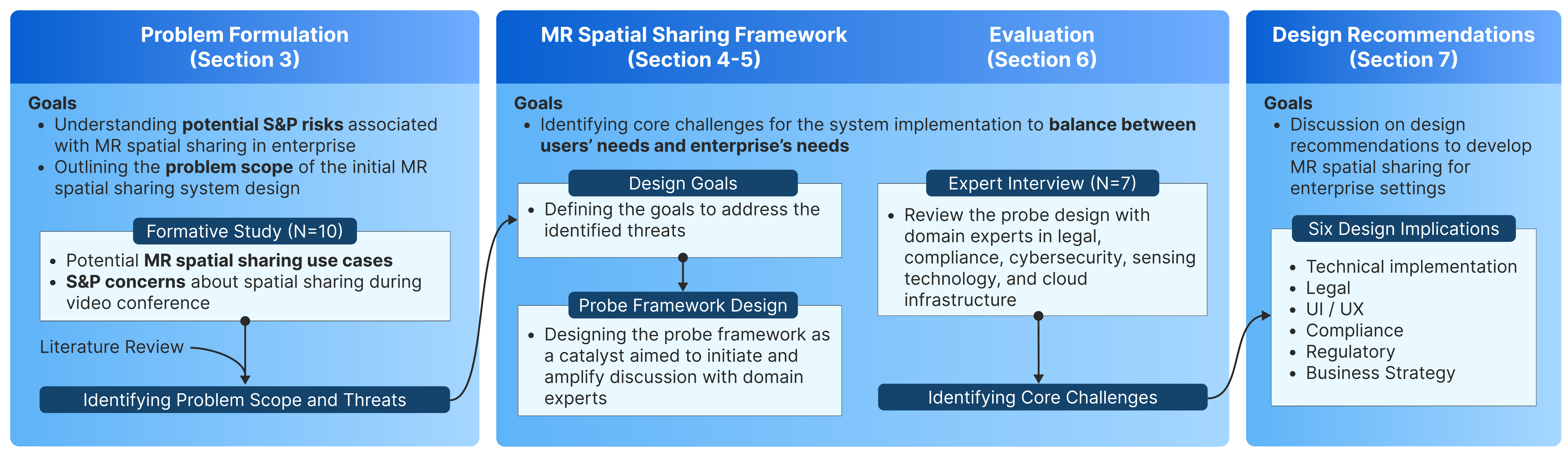}
    \caption{A diagram demonstrating the methodology of the paper.}
    \label{fig:study_structure}
\end{figure*}

\section{Problem Formulation}

%% Needing guiding paragraph

%% Formative study + Prior work ---- challenges of the current status due to lack of study in this domain (mentioned in introduction). 

During our study, we encountered two challenges to identifying the problem domains. First, there has been little research studied regarding \snp risks of MR spatial sharing in the enterprise setting. Moreover, MR adoption in enterprises is still in the early phase, and, in turn, there remains limited understanding about MR spatial sharing use cases and their contexts. 

To that end, we used two methods to formulate the problem space: (1) formative study; and (2) literature review. Both methods aimed to identify (a) potential use cases of MR spatial sharing in the enterprise contexts and (b) their associated \snp risks. Our approach was to understand the existing \snp risks in current remote collaboration practices, to see if current approaches to \snp can be transferred to MR spatial sharing, and if any potential new \snp risk could emerge. 
Specifically, the formative study explored the potential use cases employees envision and their \snp concerns during their remote work. Then, the literature review was focused on the potential use cases proposed in prior work that can be adapted to the enterprise and the studied \snp risks of remote collaboration settings.
Lastly, we consolidated the use cases and \snp risks to define our problem scope discussed in Section~\ref{subsection:problem-scope}.

\subsection{Formative Study of Spatial Sharing Practice}

% We found a lack of existing research regarding the current state of space sharing in mixed reality meeting and collaboration tools and how the users' objectives vary in an enterprise context. 

MR device manufacturers have introduced spatial sharing features and envisioned use cases in the enterprise setting, such as Meta Horizon Workrooms \cite{meta_horizon_workrooms} and Apple Vision Pro Enterprise Apps \cite{apple_vision_pro_enterprise}. However, we found two challenges in identifying \snp implications of MR spatial sharing. First, despite the increased interest in using the devices in the enterprise \cite{jalo2022extended, fortune2025xr}, they have not been well-adopted for enterprise environments due to limited established controls and policies. Additionally, we found that little has been studied regarding \snp for space capturing and sharing use cases in the enterprise context. This lack of established policies makes it difficult to derive \snp implications for MR spatial sharing use cases. 
% While it is important to acknowledge that we should not generalize it easily for different technology, we want to draw inspiration to have initial ideas regarding privacy and security concerns for XR spatial sharing by understanding the security and privacy concerns in the current remote collaboration setting on 2D screen as cameras to capture physical space are used to capture physical space both in 2D and XR. 
While we recognize that \snp considerations may differ across technologies, we want to draw preliminary insights for MR spatial sharing by examining \snp concerns in current 2D remote collaboration settings. Since both 2D and MR environments involve the use of cameras to capture and share physical spaces, these existing practices provide a useful starting point, though we remain cautious not to overgeneralize their implications to MR contexts.

Based on that idea, we conducted a formative interview study to help us understand:1) what participants currently consider as \snp concerns when sharing views of their physical space or objects during remote collaborations, and 2) how they would want to use spatial sharing using MR devices at work \cite{chen2025understanding}.
% Therefore, we conducted a formative interview study to help us understand (1) how company employees having different roles currently would practice space sharing, (2) their motivation of sharing based on their job functions as well as their privacy and security concerns. and (3) how they would use physical space sharing using XR devices.

\subsubsection{Methods}
We conducted semi-structured interviews aimed at understanding remote collaboration scenarios within a large global organization, from individuals working in different countries and representing various job functions. We recruited 10 participants via a prescreening survey who had work-from-home experience or a keen interest in using MR for work-related purposes. Each participant consented to participate in the study. Each interview was designed to last 45 minutes, ensuring sufficient time to explore the participants' experiences and perceptions in a comprehensive manner. Table \ref{tab:participant_job_functions} lists the participants' background in different areas.

The interview had three phases. In Phase One, participants shared experiences on remote collaboration and were specifically asked about their experience with physical space sharing. In Phase Two, the conversation shifted towards discussing their awareness of, and concerns about, privacy features or issues in their current remote collaborations, including their home office settings. Finally, in Phase Three, we invited the participants to envision the future of MR in the workplace. We showed participants a part of Apple Vision Pro’s commercial trailer video showing two remote users as if located in the same personal home offices. The purpose of showing the video was to stimulate discussion on expectations regarding physical space sharing (e.g., the type of content suitable for sharing, what should remain private, anticipated benefits of enhanced remote collaboration, and key privacy considerations.) The protocol was reviewed and approved by our organizational review board. 

\begin{table}[ht]
\centering
\caption{Participants' professional background for formative study}

\begin{tabularx}{\columnwidth}{l|X}
\hline
\textbf{Participant} & \textbf{Professional Background} \\ \hline
F1 & Software Development \\ \hline
F2 & Software Development \\ \hline
F3 & Software Development \\ \hline
F4 & Real Estate Finance \\ \hline
F5 & User Research \\ \hline
F6 & UI/UX Design \\ \hline
F7 & Customer Support \\ \hline
F8 & Risk Management \\ \hline
F9 & Communication \\ \hline
F10 & Software Development \\ \hline
\end{tabularx}
\label{tab:participant_job_functions}
\end{table}

\subsubsection{Preliminary Findings}
Our participants shared information about their typical behaviors during in-person and remote meetings, their needs for sharing a physical space with teammates, their privacy concerns over
displaying and sharing their space, and lastly their opinions regarding how MR could support in general their daily remote meetings and collaboration.

% \subsubsection{Behaviors of Collaboration and Content Sharing}
% Participants reported a range of behaviors and challenges in both in-person and remote settings, with several nuances emerging specifically in remote collaboration. For instance, remote screen sharing often involves dynamic role switch during situations, such as switching between pair programming, document editing, and design reviews, where real-time, instant collaboration is crucial. Moreover, participants described the need to switch between multiple applications during live presentations or demonstrations, which can interrupt the collaborative flow. In contrast, there was a strong desire for physical whiteboard sessions that support rapid and clear brainstorming through natural, face-to-face interaction. Similarly, participants expressed expectations for a virtual collaboration experience that mirrors real office interactions, emphasizing the importance of nuanced non-verbal cues. The potential of MR technologies to render near life-size representations of colleagues was highlighted as a promising avenue, as it could foster more natural interaction and focused content sharing (e.g., through features that enable users to highlight specific screen areas for detailed discussion).

\textbf{Needs for Physical Space Sharing: }
Participants frequently highlighted the appeal of spontaneous digital gatherings, such as ``virtual meeting spaces,'' which simulate face-to-face natural interactions that are otherwise diminished in remote settings. P1 mentioned: \textit{``I think being able to to quickly connect face-to-face ... if my team had a designated like clubhouse, team members could pop in every time he's in between calls.''} While digital tools like screen-based pen drawings are common for idea communication, many users nevertheless expressed a strong preference for physical media, such as whiteboards and papers for interactive brainstorming sessions, underscoring the value of tangible and analog elements in facilitating such meetings. P5 mentioned: \textit{``when you need people drawing something out on a white board, then it definitely is better to be in the office and then we will get a whiteboard and then everyone starts drawing.''} Additionally, P4 highlighted frustration of using digital tools for discussion in their work-from-home setup where there is no whiteboard , stating, \textit{"I talk with my hands. I sometimes grab a marker right on a whiteboard.... When you're working from home,... you're stuck with the tools that you got with these two screens and the keyboard and the mouse."} Moreover, P4 envisioned the benefit of using MR to address this frustration: \textit{"AR (Augmented Reality), it would be great to show someone something visually by, you know, drawing it in the air in front of me."}

There was also an expressed need for managing practical aspects. The current friction in holding webcams in hands to better capture physical objects directly impairs the effectiveness of the shared content and space. One participant showed their interest in MR as an intriguing solution to bridge this gap by enabling a more seamless and dynamic sharing environment. P3 highlighted: \textit{``currently with like the stationary camera on top of a monitor, um you can't see anything really... I think something like [MR] would be cool where you can see like the surroundings of a person.''} Its potential to combine the benefits of physical interaction and digital fluidity enhance both the clarity and interactivity of content sharing. Additionally, it was mentioned that displaying tidy and uncluttered space could contribute to a more professional image during remote customer meetings, thereby fostering a positive environment for both collaborative engagement and social interaction. P4 mentioned \textit{``I just make sure my background is clear and clean, you know, I usually tidy up if I have to before my call and that's it.''}

\textbf{Privacy Concerns for Physical Space Sharing:} 
Participants shared about their general concerns regarding the management of physical privacy in remote settings. They highlighted the importance of having \snp features for audio and video streaming as a critical control mechanism to protect personal privacy, as this allows users to rapidly restrict unintentional exposure of their environments, especially when the space is dynamic and shared with other bystanders. As P9 mentioned, \textit{``I [use screen blurring on Zoom] when I'm working from home so that other people aren't showing in the background.''} In addition to the use of background blur during video calls, a participant expressed their preference on controlling hardware settings, such as employing a hard mute or adjusting camera angles, thereby limiting the disclosure of private areas. For example, P1 mentioned \textit{``Like I should be able to just mute [my headphone] immediately ... to be able to do the same thing with the camera or my presence, you know, drop in, drop out, always connected but not always present, ... [the device]] should have like an on-device hard mute.''} Moreover, participants emphasized the need for selectively sharing portions of their physical space. P6 mentioned \textit{``for your personal preference, what do you want to show and what do you not want to show ... I'm definitely happy uh I can show my whiteboard and painting, but not the messy bed.''} P3 added \textit{``it [could] has a feature for you to draw a region of the camera's view that's going to be blacked out.''} Several users opted for sharing only curated screen contents, and suggested that smaller items would be more manageable to be readily shared, while larger, more complex physical backgrounds would need more restricted methods. Dynamic blocking of background regions, analogous to the operational principles of security cameras, was also mentioned as an effective method for safeguarding personal details: \textit{``I don't know if there is a way where, like the software that you're using as a plugin to Zoom, could detect something  privacy infringing [on] someone else's monitor.''}

%% ================= End of formative study ===============

\subsection{Problem Scope}
\label{subsection:problem-scope}

Based upon our formative study, we formulate the problem scope for enterprise remote multi-user spatial sharing in MR. To contextualize the scope of our investigation, we clarify key dimensions such as device platform, system security scope, target environments, users, data type and persistence, usage scenarios, threats and risk impact. We will illustrate each of these aspects within the boundaries of our defined scope and explicitly state the assumptions and limitations that guide our problem space definition. This approach enables a focused examination of \snp challenges specific to MR spatial sharing in enterprise practice, while acknowledging the constraints and conditions under which our findings are applicable.

% According to the insights shared by our participants, the needs and concerns for physical space sharing mainly center around the following aspects: 

\subsubsection{Case Study Applications}
% Taking these challenges into account, we identified two main axes based on needs for MR space sharing in enterprise settings as follows. The first key axis is \textbf{enterprise-level control}: are there any enterprise-level guidance or control over how to practice space sharing? The second key axis is \textbf{policy override}: does a user need to create their own rules of space sharing or does a user's preferred practice of space sharing need to override a pre-defined sharing method? Based upon these, We mapped along these two axes to help us formulate a wholistic view of the MR space sharing design space

We consider two types of workspaces in this work: the physical office and work-from-home. As \snp are context-dependent \cite{naeini2017privacy}, we consider unique \snp needs for each workspace type and aim to address sometimes different and sometimes conflicting needs of the employee and the enterprise. 
Based on the aforementioned problem formulation, our core interest is in the two scenarios below. 

\textbf{Scenario 1: Enterprise Remote Collaboration in Shared Spaces (Employee VS. Enterprise in Office)} - 
This application illustrates work settings where remote teams interact in a virtualized version of a shared office environment. The users are employees who need to collaborate seamlessly in real time while preserving sensitive elements of their physical surroundings. The primary need here is to balance immersive collaboration with rigorous security, ensuring that any background captured during virtual meetings does not inadvertently expose confidential materials or reveal employee identities. In this model, organizational IT and security teams would enforce global privacy protocols that automatically mask or blur sensitive visual data, thus removing the risk of user misconfiguration while delivering a consistent, secure experience across the organization.

% \textbf{Scenario 2: Remote Real Estate Inspection} - there is a need for a mixed reality tool tailored to property evaluation processes used by real estate professionals. The envisioned users are property inspectors, real estate agents, or potential buyers who benefit from an immersive view of facilities that combines both physical and digital insights. In this case, the technology would allow for dynamic presentation of a property, offering the option to customize which parts of the environment are visible or concealed. The system would implement enterprise-level privacy settings to protect confidential information, whether it involves proprietary layouts, client data, or other sensitive details, providing a controlled experience that supports decision-making while mitigating privacy risks.

\textbf{Scenario 2: User-Controlled Environment Sharing for Home Offices  (Employee VS Enterprise for Work from Home)} - 
Another application centers on the needs of remote workers sharing live views of their home offices. In this case, individuals seek full control over what they disclose in their personal spaces during virtual interactions. This user-centric approach emphasizes customizable privacy settings, allowing each person to decide on a case-by-case basis what elements of their environment are shown or hidden. The flexibility here meets an important need for personal privacy management while fostering a sense of openness and connectivity among remote teams. Nonetheless, such designs also come at the cost of potential privacy leakage due to unapplied corporate enforcement and may introduce specific security challenges, especially when considering targeted threats.

\subsubsection{Device Platform and Diversity}
% \todo{need to review this part of device configuration, does this align with our scope?}

In this study, we consider MR devices as a general category featuring external cameras available for spatial sharing, without restricting our scope to a specific platform or manufacturer. Examples of MR devices include enterprise-managed solutions such as Apple Vision Pro, Meta Quest, Samsung Galaxy XR, and other custom business-oriented platforms. Our focus is on devices that are managed by the enterprise for business use, ensuring that appropriate \snp policies and controls can be enforced.

We also acknowledge scenarios where participants may use personal MR devices, such as when an employee meets with an external client who connects using their own hardware. In these cases, at least one device in the session is assumed to be enterprise-managed, and the MR spatial sharing application itself remains under enterprise management and control, even when installed on a personal device. This approach allows us to address a broad range of device configurations and security capabilities relevant to enterprise MR spatial sharing.

\subsubsection{System and Communication Security}

For the precise positioning of the problem space we want to tackle, we define the setup for a system that we target to build as follows. We assume that data communication is secured and cannot be bypassed by unauthorized entities. Thus, all data is accessible only when a user activates data sharing, meaning that the problem space scope is within risk caused by a user's actions during MR spatial sharing interactions.

% \todo{please review the following statement on app security}

Furthermore, as the MR collaboration application itself is managed by the enterprise, we assume that the application and its underlying infrastructure adhere to enterprise security standards and are protected against unauthorized access or tampering. This allows us to focus our analysis on user-driven \snp risks within the MR spatial sharing context.

\subsubsection{Target Environments \& Associated Items}

Our focus for environment is two-fold: a physical office space and personal space for remote work. For a physical office, enterprises could have their own organization-wide \snp policies and compliance requirements for their employees to adhere to in terms of spatial sharing. Specifically, there are three key items involved in a physical office environment that may cause \snp concerns in a spatial sharing session: 1) physical objects, 2) physical space, and 3) other co-located employees. For example, a physical space may have objects such as documents or monitors on other desks that have companies' sensitive information. Additionally, spaces could have highly restricted access and disallow any information to be visible from outside. Lastly, employees often share a physical space with other colleagues, which may lead to potential recording of other colleagues without their knowledge.  

For personal space, we envision a setup where an employee works at their desk and there are co-located personal items or people such as family members that are not supposed to be shown to unauthorized entities through spatial sharing. 

\subsubsection{Target Users}
% \todo{please review the following statement on target users and their consent}

Our target users are enterprise employees who participate in MR spatial sharing, either in office or remote work settings. These users may occupy various roles and work in environments with differing privacy requirements. We assume that users of MR technologies have provided consent for their physical space to be captured by the device as a prerequisite for using the hardware. Furthermore, users are aware that, during MR spatial sharing sessions, their visual environment may be streamed or shared with other participants via the cameras, similar to the expectations set by conventional video conferencing tools. Our scope centers on users who understand and accept these conditions as part of their participation in MR spatial sharing.

\subsubsection{Data Type and Persistence}
% \todo{please review the data persistence part}

In this paper, we focus on visual data as the data type to manage. 
Other data types also potentially have companies' sensitive data such as business-related conversations in a physical space, which could be unintentionally recorded in another employee's video conference. Audio data, for example, would need different filtering methods to protect data from visual data. Visual data can be invisible by simply covering the data from cameras while covering a sound source cannot guarantee inaudibility as sound could propagate through physical media. Therefore, addressing risks associated with all the other modalities is out of scope for this paper. While acknowledging risk from other data types, we limit our scope to visual data as a first step to address due to it being a dominant channel of how users engage with MR experience.

We also assume that visual data is stored and processed only for the duration necessary to support active MR spatial sharing sessions, and is not retained beyond the session unless explicitly authorized or required by enterprise policy. Persistent storage, archival, or secondary use of visual data is considered out of scope for this study. Our analysis focuses on the \snp implications of visual data during live interactions, under the assumption that appropriate data deletion and retention policies are enforced.

\subsubsection{Identified Threats}

The mixed reality environment for the above use cases poses the risk by a malicious insider in a virtual meeting. Specifically, an adversary who participates in the same virtual environment as a legitimate user could deliberately record sensitive parts of the user’s background in order to extract confidential information. Potential targets include sensitive documents, personal identifiable information inadvertently captured in the background, or even details about the physical layout of either a private space or a highly restricted area. We exclude potential threat where a malicious actor has access to a physical space of a user where target objects are located. 
% In scenarios where both participants are equipped with head-mounted displays and external sensors, ensuring that sensitive information is adequately protected requires careful system design. The assumption here is that while the underlying conference software adheres to strong security standards (and is thus immune to direct manipulation), the interface design must still provide robust measures, a combination of organizational policy enforcement and user-specific customization, to mitigate the risk of covert data collection by a malicious actor.

% \subsubsection{Risk Impact}
% \todo{Please review the risk impact, consequences of identified threats}
\snp failures in MR spatial sharing can have serious consequences for enterprises. From the enterprise's point of view, data leakage, such as the unauthorized exposure of sensitive documents, personal information, or confidential business operations, can result in financial loss, regulatory penalties, and legal liabilities. Moreover, reputational harm may arise if clients, partners, or the public lose trust in the organization’s ability to safeguard information. From the employee's point of view, unwanted recording of personal space through MR technologies can 
engender their \snp concerns regarding their personal space.

Based upon our formative study, we formulate the problem space for enterprise remote multi-user spatial sharing in MR. To contextualize the scope of our investigation, we clarify key dimensions such as device platform, system security scope, target environments, users, data type and persistence, usage scenarios, threats and risk impact. We will illustrate each of these aspects within the boundaries of our defined scope and explicitly state the assumptions and limitations that guide our problem space definition. This approach enables a focused examination of \snp challenges specific to MR spatial sharing in enterprise practice, while acknowledging the constraints and conditions under which our findings are applicable.

% According to the insights shared by our participants, the needs and concerns for physical space sharing mainly center around the following aspects: 
\section{Design Consideration}
\label{section:designconsiderations}
% Taking these into account, we identified two main axes based on the use cases of MR space sharing in enterprise settings as follows. The first key axis is \textbf{enterprise-level control}: are there any enterprise-level control over how to practice space sharing? The second key axis is \textbf{policy override}: does a user need to create their own rules of space sharing or does a user's preferred practice of space sharing need to override a pre-defined sharing method? Based upon these two axes and current literature, we consider four use cases mapped along these two axes to help us formulate a wholistic view of the MR space sharing design space. From these selected use cases, we aim to derive our security and privacy goals.
Following our identified \snp challenges, usage scenarios and possible threats, we develop the following functionality and security goals with an aim to construct a safe while usable MR space sharing framework for enterprise environment.  

\subsection{\snp Goals}

MR spatial sharing in the enterprise setting is associated with a unique context for \snp where there are \snp demands from both employees and enterprise. In addition, as the remote work setup has become more ubiquitous than before, employees could work in another physical space (e.g., a personal space for work from home), which may require different \snp needs from the office space. Therefore, our \snp goals are tied to understanding how to account for the needs of employees and enterprise, how to balance them, and how to apply \snp requirements for various physical space contexts for MR spatial sharing. 

\begin{itemize}

    \item \textbf{Content for Filtering.} In this work, we have two types of content of interest for filtering: physical objects/space and biometric data of bystanders. First, camera capture of physical objects and space in the office space during a call can unintentionally disclose sensitive information. For example, a user may take screenshot of sensitive document placed on an office desk and identify confidential information. Likewise, images captured of a personal space during remote work could help infer sensitive information about a user \cite{prange2022saw} (e.g., inferring financial status based on visible items). Second, cameras could record undesirable facial and other biometric data. Ubiquity of cameras could increase the unwanted recording of bystanders and, in turn, raise their \snp concerns \cite{denning2014situ,yao2019privacy,prange2022saw}. In the context of MR spatial sharing, a spatial sharing session may include employees' faces accidentally appearing in the background of the main subject or in an area of focus and this may in turn leak employee personally identifying information (PII)  to unintended audiences or non-employees. Similarly, MR spatial sharing could capture bystanders' facial data in the employee's personal space such as family members who employees may not want to show during a business call. 

    \item \textbf{\snp Verification of Enterprise Policy.} Protecting internal data is a primary concern for a company \cite{}, leading them to require employees to take care that sensitive data is not unwittingly being shared outside the firm. To that end, one of our design considerations is that information regarding users' spatial sharing is verified by the enterprise's spatial sharing policy before streaming. The could be cases where employees may want to override the enterprise's default \snp settings. For example, during work from home, employees may want to turn off the default background blur feature to show a whiteboard in their personal space. Even in this case, the employees' desired settings need to go conform to the enterprise policies and regulations to mitigate the risk of unintended disclosure of the company's sensitive data.

    Additionally, we consider that enterprise access control seeks to ensure that only people of the right access could see physical objects during the spatial sharing session. We draw inspiration from existing collaboration tools' features that allow a project owner to control which person can access which file (e.g., Microsoft Teams' access control). For example, while enterprise policy may have no restriction regarding sharing physical objects for employee-to-employee meetings, the policy could be applied to filter the camera capture of documentation or other monitors co-located in the employee's physical space during a meeting with clients from another firm. 
    
    \item \textbf{Employees' Spatial Sharing from Different Locations.}
    Enterprise's \snp policy could be misaligned with employees' \snp expectations depending on locations. Specifically, the enterprise regulation and policy may not necessarily need to apply to users' personal space. For instance, employees may accidentally show personal documents that include sensitive personal information and this information may not be filtered by the enterprise policy and regulations because it is not related to the enterprise's sensitive information. In this case, employees may want to apply additional \snp controls to secure their own \snp. 

    Similarly, employees may want access to \snp control adjustment even in the physical office spaces. Different parts of the physical spaces may not need the same level of \snp control. Places where highly confidential information is discussed such as certain executive spaces would need strict policy and regulations regarding spatial sharing while employees may want to show whiteboards or everyday physical items in spaces like general conference rooms, not needing the strict regulations for spatial sharing. Thus, it is important to understand the devices' location during spatial sharing, allowing employees to have different levels of agency over \snp control. To do so, device localization techniques would need to be leveraged.
\end{itemize}

\subsection{Functionality Goals}
Based on the usage scenarios we identified, we derive a set of functionality design goals for MR spatial sharing applications and platforms in enterprise settings. These functionality goals related to spatial sharing can help ensure the usability of multi-user mixed reality apps when sharing functionality is required.

\begin{itemize}
    \item \textbf{Support contextual awareness} - For all of the usage scenarios, knowing where the user is physically located and what physical objects exist in the current space is crucial to helping both the user and the system determine the best privacy-preserving strategies for spatial sharing \cite{slocum2024}. A multi-user MR platform would need to support localization and scene understanding (for the physical space around the user) before initiating spatial sharing. 
    \item \textbf{Adaptive Space Filtering} - For home office collaboration, the user would need to selectively share their home space based on certain requirements, such as filtering out a certain area in the room or certain objects that are visually sensitive~\cite{kim2023erebus, sidenmark2024}. Enabling adaptive space filtering could help provide more fine-grained control over how the space is to be shared.
    \item \textbf{Support real-time and immediate control} - As the spatial sharing practice depends heavily on the physical situation where the user is located and this situation could be highly dynamic, it would be important to enable real-time and immediate control over the spatial sharing settings (e.g., turn on or off the sharing) to make sure the control changes can be applied without any delay when needed \cite{Rajaram2025, rajaram2023eliciting}.
\end{itemize}

\subsection{Non-Goals}
We consider the following design questions to be non-goals of our investigation:

\begin{itemize}
    \item \textbf{Spatial sharing control options.} We recognize that each user or organization must determine what information should be shared or withheld based on their unique use cases and requirements. Our work provides example categories and types that are commonly associated with privacy concerns, but it does not prescribe specific control options or configurations.
    \item \textbf{Advance scene understanding and object segmentation techniques.} While crucial for enhancing object-level recognition and tracking, it is not the primary focus of this paper. Although such capabilities are important for improving the accuracy of spatial sharing and functionality of MR applications, our work concentrates on addressing \snp concerns within the space-sharing framework. 
    \item \textbf{MR system implementation and user interface design.} Our design operates at a lower level so it does not aim to define how the experience should manifest to the users. We do not aim to enhance user performance in practicing spatial privacy through interface design or implement a working system; rather, we focus on providing a robust framework for safeguarding \snp in shared spaces, empowering enterprise users to implement privacy-preserving space-sharing principles and techniques.
\end{itemize}

\begin{figure*}
    \centering
    \includegraphics[width=\textwidth]{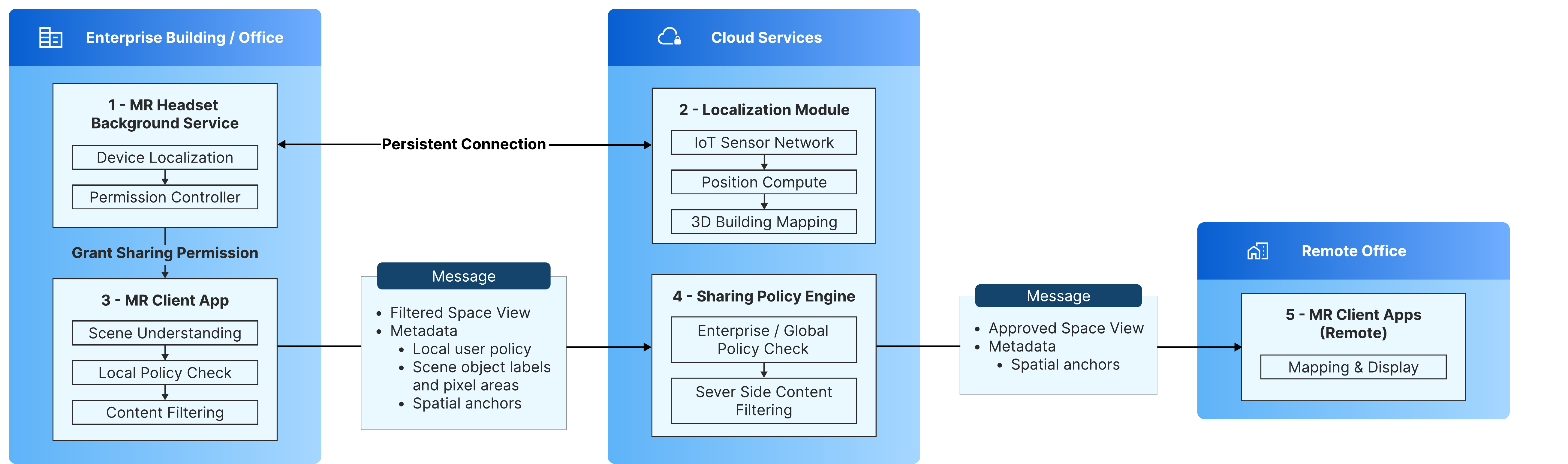}
    \caption{Our conceptual framework adopts a server-client architecture and the spatial data passes through the major components in the following order: 1) MR background service, 2) cloud localization module, 3) MR client app, 4) cloud sharing policy engine, 5) remote MR client app.}
    \label{fig:system_diagram}
\end{figure*}

\section{Study Design}

%  The protocol was reviewed and approved by our organizational review board. 
The challenges and problem space identified above for MR spatial sharing in enterprise settings are multi-faceted. Our design considerations encompass key aspects of usability, security, and technology, all of which can significantly impact the performance of employee \snp protection. Given these complexities, we sought to further study how to properly design a spatial sharing system, that well aligns with existing enterprise \snp practices, as well as assessing the trade-offs between user utility and enterprise \snp requirements. 

Due to the lack of widely adopted enterprise MR platforms and the limited exposure or knowledge most enterprise users have with such systems, assessing the design requirements through traditional means is challenging. To address this, we first designed a prototype framework and spatial sharing system that can serve as a `technical probe', or a hands-on technical instrument, to instigate conversation and discussion that we could get feedback for. Our intention is to show a brief idea of the system and to ``provoke inspirational responses'' from the domain expert \cite{gaver1999design}. Then we use that prototype framework in a domain-expert interview study and invited the experts to comprehensively envision a possible real-world deployment and provide feedback that departs from the probe system. Our study had two primary objectives: 1) to understand what domain experts perceive as the primary challenges and potential design solutions when integrating MR spatial sharing within the existing enterprise \snp frameworks, and 2) to gather insights on how experts would design a comprehensive MR spatial sharing framework that minimizes friction during actual deployment. We will discuss how to design and develop the probe in the following subsections.

% Before building the entire system, we asked domain experts in different areas that the system covers about key factors to consider for the implementation. To do so, we used `technical probe' where we designed and developed the minimal function of the system. Our intention is to show a brief idea of the system and to ``provoke inspirational responses'' for the domain expert \cite{gaver1999design}, helping us collect the major factors at the early stage without re-building the system. 
% We will discuss how to design and develop the probe in the following subsections. 

% \subsection{Probe Design}

\subsection{Probe Framework Design}

% We envision our system to consist of a head-mounted display device connected with a server and expect the device's feature to be run based on the policy determined by the relationship between the local device and the server. In the following sections, we will describe our initial framework to design our probe. 

% previously design section below
Following design considerations listed in \Cref{section:designconsiderations}, we now present a proof-of-concept MR spatial sharing system based on high-level conceptual framework that enterprise architects may consider when designing their own system to enable secure and private space sharing in mixed reality. We envision a system of this sort would contain both app-level libraries and OS-level background services along with cloud services for enterprise-wide deployment.

\subsubsection{Spatial Sharing Components Overview}
To support a secure and private multi-user MR spatial sharing experience, we provide an overview of each component in the spatial sharing workflow. Figure \ref{fig:system_diagram} is a framework diagram of the key steps of sharing from outbound to inbound between two clients and the key space data processing units. The proposed system architecture for privacy-preserving spatial sharing in mixed reality (MR) environments is comprised of five core components, each responsible for distinct functions within the workflow. The architecture is designed to enforce \snp through layered policy checks and content filtering, both locally and globally at the enterprise level. Below we illustrate how the main components of our framework work together:

\textit{MR Background Service} - This service operates continuously on the MR device, initiating localization requests when the device is powered on. It is responsible for establishing secure communication with the cloud-based localization module to verify permissions for spatial sharing based on device location and enterprise access controls. By enforcing location-based access restrictions, the background service ensures that spatial sharing is only permitted in authorized areas.

\textit{Cloud-based Localization Module} - This module operates in the cloud and manages a network of IoT sensors (e.g., WiFi routers, ultra-wideband sensors, Bluetooth beacons) distributed throughout the enterprise environment. Upon establishing a persistent connection with the MR device, the module aggregates sensor data to accurately compute the device’s spatial position, such as by triangulating its distance from multiple access points referenced against a digital 3D building map. The computed location is sent back to the MR background service and simultaneously serves as a key permission condition for enabling subsequent spatial sharing functionalities. The motivation for centralized localization is to minimize reliance on device-side computer vision, which often yields limited accuracy in large enterprise environments, while also enabling the system to establish buffer zones that proactively restrict access as the device approaches non-permitted areas.

\textit{MR Client Application} - The MR app is the primary user interface for remote spatial sharing. After receiving permission to share the space from the background service, the app could then perform scene understanding to identify and label physical objects within the environment. It applies user-defined local policies to determine which areas and objects are eligible for sharing. Content filtering is performed to mask or blur non-shareable elements before any data is shared. After the filtered space view is prepared, the app sends out this view in a message along with some metadata which includes user policy specification, scene object labels, and pixel area mappings for each object.

\textit{Server-side Sharing Policy Engine} - Upon receiving the locally filtered view and metadata from the MR client app, the server-side sharing policy engine performs a global policy check based on enterprise-wide \snp rules. Additional content filtering may be applied, including role-based access controls that could tailor the shared view for individual recipients. This layer provides a second, institution-wide safeguard, ensuring that all shared content complies with organizational standards and regulatory requirements.

\textit{Remote MR Client App} - The server-approved space view is transmitted to remote MR client apps. Each recipient device maps the incoming view to its local spatial context and renders within the headset for user interaction. Only content that has passed both local and global policy checks is shown, ensuring end-to-end privacy protection throughout the sharing session.

\subsubsection{Probe Framework Implementation}
\begin{figure}
    \centering
    \fbox{\includegraphics[width=\columnwidth]{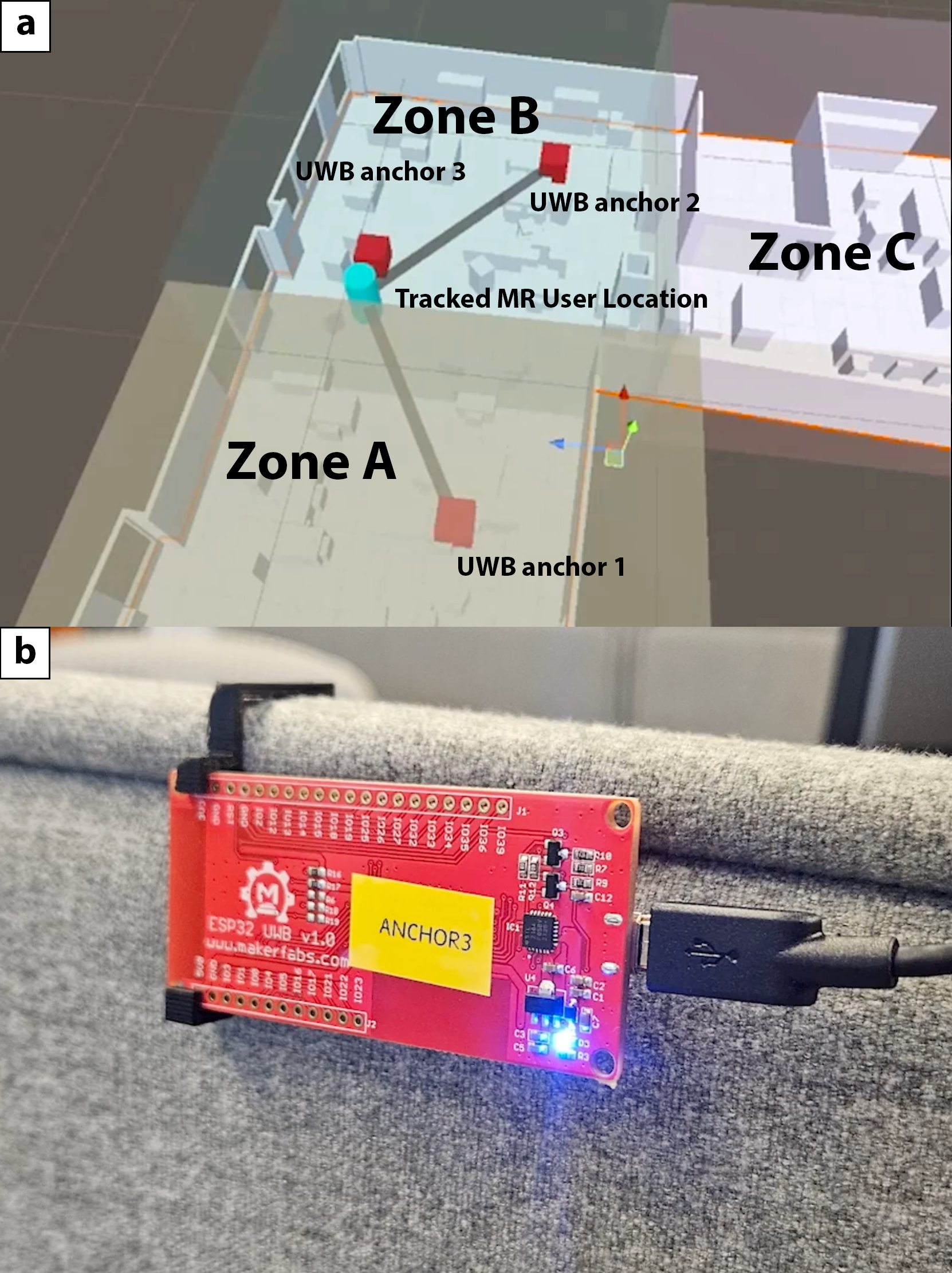}}
    \caption{Visualization of our probe application. (a) shows a visual map of real-time MR device tracking through ultra-wideband (UWB) anchors in a digital twin environment, with distinct zones highlighting area-specific access permissions. (b) shows one of the UWB hardware sensors used for MR device localization.}
    \label{fig:implementation_demo}
\end{figure}

The implementation of our probe system draws inspiration from prior research on privacy-preserving spatial sharing in mixed reality environments \cite{cheng2024spatialprivacy}. Building upon these foundations, we developed a prototype system that integrates an ultra-wideband (UWB) sensor network for localization, alongside a digital twin of an interior office space generated using the Apple RoomPlan API \cite{apple-scene}. Subsequently, we created a Swift library targeting iPadOS and VisionOS to enable location-based permission control, policy checking, and content filtering. Additionally, a Python-based global policy service was developed and deployed on an AWS EC2 instance to support global policy checks and content filtering. Figure \ref{fig:implementation_demo} presents a sample application built with the toolkit along with a live visualization of device localization in the digital twin space, demonstrating spatial sharing in accordance with the probe framework design. This implementation serves as a technical demonstration of the design for expert interviews and validates the feasibility of our conceptual architecture.

\subsection{Participants}
We recruited domain experts from a diverse range of fields, including cloud engineering, usable security \& privacy, cyber-security, ubiquitous computing, legal, and technology risk and control from a large global organization. The intent was to capture a comprehensive perspective on both the challenges and potential design solutions in the context of enterprise \snp practices. Table \ref{tab:expert_background} shows a total of 7 domain experts we invited for interviews, each with 8 to 32 years of experience in their respective fields. Each domain expert consented to participate in our study.

\subsection{Study Procedure}
Our study protocol was reviewed and approved by our organizational review board. We invited each of our participants to join a Zoom meeting to conduct a 60-minute semi-structured interview, as we aimed at comprehensively exploring our conceptual framework and gathering detailed feedback from our participants. The interview process was divided into several key phases: 1) \textit{Introduction} - to confirm expertise and discuss current major concerns regarding enterprise \snp; 2) \textit{Conceptual Framework Walkthrough} - participants were presented with artifacts (video demos, diagrams, slides) from the prototype framework and given a detailed explanation of the concepts, key design considerations, and fundamental principles of MR spatial sharing; 3) \textit{Feedback Collection }- to gather domain-specific insights on the framework's effectiveness and limitations; 4) Real-World Challenge Discovery - we deep-dived with the experts in discussing the potential challenges and blockers that could hinder the effective protection of user \snp in a real-world scenario.

% We tailored the interview questions to each participant’s expertise. They were broadly categorized into three areas:
% \begin{itemize}
%     \item \textbf{Technical \& Infrastructure} - covering questions on engineering challenges, data processing, service integration, XR device management, scalability, and algorithmic issues.
%     \item \textbf{Usable Security Design} - focusing on finding additional security risks, identifying potential threats, understanding usability trade-offs, incident response strategies, XR-specific UX design, and managing user permissions.
%     \item \textbf{Legal and Risk Control} - covering threat modeling, data governance, legal compliance, user consent, and incident reporting and management.
% \end{itemize}

\subsection{Interview Questions}
We tailored our interview questions to align with each participant’s domain expertise, ensuring that the discussions were both relevant and insightful. The questions were organized into three principal categories and used according to each participant’s area of specialization:

\setlength{\fboxsep}{0pt}

\begin{table}[ht]
\centering
\caption{Expert participant background and years of industry experience (YOE)}

\begin{tabularx}{\columnwidth}{l|X|l}
\hline
\textbf{PID} & \textbf{Area} & \textbf{YOE} \\
\hline
% P1 & Sr Engineering Lead & 35 yrs \\
% P2 & Lead Cybersecurity Architect & 8 yrs \\
% P3 & Director of Business Information & 20 yrs \\
% P4 & Director of IoT Research & 10 yrs \\
% P5 & Sr Principal Security Architect & 15 yrs \\
% P6 & Legal Risk Lead & 9 yrs \\
% P7 & Head of Risk and Compliance & 32 yrs \\
P1 & Enterprise Cloud System Engineering & 35 yrs \\
P2 & Cybersecurity Architecture & 8 yrs \\
P3 & Business Information Systems & 20 yrs \\
P4 & IoT \& Emerging Tech Research & 10 yrs \\
P5 & Cybersecurity Architecture & 15 yrs \\
P6 & Legal, Governance, \& Risk Management & 9 yrs \\
P7 & Risk \& Compliance & 32 yrs \\
\hline
\end{tabularx}
\label{tab:expert_background}
\end{table}

\begin{itemize}
\item \textbf{Technical \& Infrastructure:}
This category encompassed questions designed to elicit expert perspectives on the practical and theoretical challenges of implementing sensor-enabled and XR systems within enterprise environments. Topics included secure data transfer, system integration with existing enterprise services, device management, scalability, reliability, and the development and maintenance of both hardware and software components. Participants were also asked to reflect on algorithmic challenges such as accurate localization, scene understanding, and privacy-preserving data processing.

\item \textbf{Usable Security Design:}  
Questions in this category focused on the intersection of security and usability, probing how systems can be designed to balance robust security controls with user experience. We explored participants’ experiences with identifying and mitigating security risks, managing user permissions, and designing incident response strategies. Additional emphasis was placed on the unique challenges posed by XR environments, such as spatial sharing, automation of policy enforcement, and the detection of malicious intent, as well as the role of enterprise monitoring and feedback mechanisms.

\item \textbf{Legal and Risk Control:}  
This category addressed the regulatory and ethical dimensions of deploying sensor-enabled applications. Interview questions examined participants’ experiences with legal compliance (e.g., GDPR), data governance, user consent, and incident reporting. We also discussed the development of comprehensive threat models, the design of access control mechanisms, and the formulation of effective incident response plans. Consideration was given to the administrative process of onboarding new devices and ensuring that employees are adequately trained to adhere to \snp protocols.

\end{itemize}

Before starting the formal study, we piloted the questionnaires with three additional domain experts. Their feedback enabled us to refine and adjust the questions, ensuring they were well-aligned with the study’s objectives and maximized the relevance and clarity of each interview.

\subsection{Data Collection and Analysis}
Data for this study were collected through audio-recording of the interviews with domain experts. The recordings were transcribed using an internal transcription tool based on the AWS transcription service, which complies with the participants' institutional policies. The resulting text transcripts were then exported for qualitative analysis. As the primary goal was to capture expert perspectives on the main challenges and potential design solutions for MR spatial sharing, three researchers independently conducted thematic coding of the transcripts to identify key characteristics and recurring themes. An inductive thematic analysis approach outlined by Braun and Clarke\cite{braun2006using} was employed to systematically analyze the data and derive meaningful insights. After that, the researchers held meetings to discuss and refine the codes until reaching consensus. These codes were further analyzed and reviewed, resulting in the identification of two overarching themes and a total of nine sub-themes.

% we interviewed x number of experts, we showed things and asked questions, table of their background, yoe

\section{Results}
\label{section:results}
Based on our open coding and thematic analysis, we identified two themes: 
1) Technical, Interaction Design, and Deployment Challenges, and 2) Management Challenges. Within each theme, we further identified a total of nine subthemes, which will be discussed in the following sections.

\subsection{Technical, Interaction Design, and Deployment Challenges}
One part of our focus is to gain a holistic view of the technical and design challenges to address to make a functional and usable system. 
We will discuss the expert interviews on technical challenges and how those would impact performance, robustness, and the users' interactions with the system. Then, we will discuss the potential remaining physical deployment challenges if the technical and design challenges are addressed, including hardware infrastructure and device management. 

\subsubsection{Space Management and Localization}
Context understanding in physical space is one of the key elements for XR spatial sharing. 
In particular, XR devices allow a user to transition between different physical spaces, some of which could be permissible and others non-permissible. Therefore, a system needs to employ context-dependent features to continuously locate the users and handle the environment transitions. For example, P1 questioned \textit{``how do you manage that basically I started the connection with the other person and I suddenly moved to a sensitive area ... how do you disable as soon as you see that?''} However, a faulty contextual understanding could lead to unintended usage reduction or enablement, posing risk of exposing contents to unauthorized individuals during XR spatial sharing. For example, P4 commented \textit{``you think you're on a red [non-permitted] floor when you're on the green [permitted] floor. So you disable sort of content sharing. Maybe it's not the end of the world, right? But the converse of that situation would be pretty bad.''}

To do so, participants mainly emphasized two ways to reduce risk. First, several participants discussed the localization accuracy during a user's continuous movement (e.g., moving to another floor) and how to manage the user's space transitions between permitted and non-permitted areas in the office. P2 mentioned \textit{``perhaps you could augment those headsets with RFID readers, which employees can tap their badges on ... if they go from our unrestricted space to a restricted space, I think that should be made clear to an employee.''} Additionally, several participants recommended the default settings to mitigate the risk by adopting ``whitelisting'' where every items are blocked by default except items selected to be allowed, instead of ``blacklisting'' strategy where every items are allowed by default unless items are explicitly selected to be disallowed. In doing so, when the system's localization works inaccurately by accident, the whitelist strategy still blocks the content that was allowed to be shown while the blacklist strategy would expose the contents that are disallowed. P5 emphasized the need of adopting the whitelisting strategy by saying \textit{``it's an allow list versus a deny list ... an organization can apply a policy that says no more maximum than this. So ... if somebody accidentally has an allow list all for global, that then doesn't interfere with something else.''}. Similarly, P6 echoed by saying \textit{``By default, everything is turned off and then you would have to request an exemption to enable the turning on of the feature. Because that's a little bit like how we handle data loss prevention within the firm.'}'. 

\subsubsection{Performance and Optimization}
Many participants expressed their concern on the ramification caused by system failure, leading to security breaches, and urged the need of robust system performance. For instance, participants mentioned how the system's latency could affect security. Specifically, if there is latency on blurring a sensitive content due to blurring processing or data communication, the content would be unintentionally disclosed causing security breaches. P4 mentioned \textit{``if there's any sort of a noticeable latency involved, either continuous or maybe due to some burst ... that could potentially severely degrade the whole experience, which might or might not be the make-or-break point.''} P2 echoed by saying \textit{``if there's like a severe delay in server side filtering, it might lead to a pretty unuseful experience, on top of perhaps like sensitive things getting leaked for a brief moment.''} To mitigate this potential risk, participants also made suggestions on the system design. One way to do so is to reduce the amount of data processing for blurring as the more data to process the more time to process. For example, downsampling the camera feed or pruning redundant image processing could help optimize data processing and, in turn, reduce the latency. \textit{``let's say the first frame is completely processed in terms of going through your entire pipeline. but the second frame is not that much different from the first, then probably a lot of the processing can be omitted, but still resulting in the same output in terms of which region is to be blocked (P4).''}

Additionally, participants emphasized the benefits of running local processing for computer vision process. To be specific, the image processing on local machine that runs the policy control enforced by enterprise negates the need of continuous server connectivity, reducing the time taken during the server-to-device communication and, in turn, lowering the latency. P5 mentioned \textit{``I believe doing a triangulation computation locally on a simplified version of the policy ... would not be computationally expensive ... the idea being that you are already having to compute your position within 3D space in the first place.''} Furthermore, the local processing would be resilient against unexpected server disconnection, safeguarding the computer vision process to run and eroding the risk of the unexpected exposure, which in return could lead to an improved user experience \textit{``any persistent connection that independent of potential security problems of like what happens if a disconnects gets spoofed and all that, it's a better user experience in general (P5).''}

\subsubsection{Awareness Improvement on Policy Control}
Once the technical challenges are addressed, participants shared suggestions from the end-user's perspectives regarding how to minimize user errors and enhance policy awareness.
In particular, participants stressed the importance to ensure that users have a good understanding of policy settings to avoid confusion about what policy controls exist and how they operate and are adjusted subsequent context changes. As P2 elaborated: \textit{``are users made aware of the enterprise policies for what gets filtered out? Because maybe it would be their intention to perhaps share some documents and it could get mislabeled as sensitive.''} The wrong understanding could create the gap between how employees expect policy controls to operate and how they actually operate and make users unable to perform spatial sharing for what they want to demonstrate during the call (e.g., an employee wants to clear the policy regulation around the contents that they want to show yet not approved by default.) To close the gap, experts placed emphasis on the in-advance communication about the policy settings, which would further consolidate intention groundings and avoid delays in business: \textit{``users definitely should be made aware of the default setting before you start doing any sharing. And if there are any changes, that should be communicated to them very well in advance. (P2);''} and \textit{``I think it could present a pretty big usability hurdle ... if you're already on a meeting with a client and you wanted to share a document and then that's filtered out [by the headset], that could potentially annoy a client from their perspective (P2).''}

While the policy communication prior to the meeting is suggested, participants also discussed the importance to communicate the policy on-the-fly. To be specific, visual warning when approaching sensitive area would help employees to be reminded and understand how the policy works in real time. In doing so, users could be cautious about their spatial sharing and reduce risk of any data breach by moving back to the pre-cleared area: \textit{``are there any warnings that happen if you could sense someone approaching a restricted space or leaving an unrestricted space? ... I think definitely keeping a user informed. it's a big thing (P2).''}

\subsubsection{User Agency over Policy Control}
In the previous subsections, we presented participants' discussion on how enterprise needs to communicate their policy control with employees. Experts raised the importance to enforce organizational policy control to mitigate risks.  P5 mentioned \textit{``a good example of this that you can look up later is how AWS IAM policies apply ... They start with the strictest definition. And then you can only allowlist in additional things, except for there are certain conditions where even if a more looser policy gets applied, an organization can apply a policy that says no more maximum than this. So ... if somebody accidentally has an allow list all for global, that then doesn't interfere with something else.''} 

On the other hand, participants discussed considerations regarding how users can have agency over the policy control mainly for two contexts. First, in practice, object recognition's accuracy will not be ``perfect'' where non-sensitive contents can be misclassified as sensitive and obfuscated due to the system failure. P2 mentioned \textit{``Have you considered giving any leeway to the users who perhaps override or request an override of the content filtering?''} P5 mentioned \textit{``you start off when you first boot, maybe it's not as ideal because it [the algorithm] does filter things by default, or if it doesn't understand it, then maybe it's going to fail close, but by doing that content filtering as soon as possible, you're just taking out of the equation...it gives more privileges or permissions to the user to control how their environment behaves. So it's in the same way that when you share on Zoom.''}

However, the opposite case where sensitive contents are classified as insensitive would pose higher risk as it leads to the sensitive material exposure: \textit{``you're on red, but you think you're on green, then you start sharing something on or maybe, you know, from a scene understanding, there's this computer screen, but let's say the CV didn't catch that ... then that's a big risk, to say the least (P4).''} Therefore, a user's agency over the enterprise control to correct this misclassification could prevent the sensitive material from being accidentally exposed to unauthorized people in the call (e.g., mute button for HMD's external cameras to disable any video capturing): \textit{``another thing that I was thinking about was ... something similar to a mute or a hide camera button in zoom, like giving employees the ability to do that would be would be pretty nice (P2).''}

In summary, providing users with agency regarding policy overriding could benefit not just usability of the system but also security of the system. However, this is context-dependent according to the aforementioned examples. 
% From enterprise perspective, their goal is to prevent any security and privacy breach. They often do so by enforcing policy control for employees to follow and regulate their actions. However, their preventive enforcement in return erode users' agency over the system control, potentially reducing the usability. As security and privacy are secondary concerns for typical users \cite{}, users potentially would find a way to detour from the secure way to complete the task if motivated to finish their work as soon as possible, which leads to security and privacy breach \cite{}. This case applies not just to employees but also to client experience. 
\subsubsection{Business Needs and Costs for New System Development and Setup}

Deploying a new hardware system in the physical office space could encounter various practical challenges tied to laborious and financial burden. Some participants mentioned that the deployment would require balancing business needs and costs: \textit{``you also have to think about the costs that would incur for the firm (P2);''} and \textit{``when it comes to the financial side, the business value side, that needs to be argued then presented ... from the business value from the money side, that needs to be a convincing case for them (P4).''}

As aforementioned, handling various cases of XR spatial sharing appropriately is important and inevitable, leading to the needs for a complex system design. This complexity could give huge burden to developers and discourage them from actually integrating the framework to the internal systems at the organization. P5 pointed out developers’ willingness to learn and adopt the framework. P5 expressed concerns that if the policy is complex and hard to learn and build, it would cause the friction for developers to implement the `secure' framework: \textit{``you need to be able to express to the developer why they should adopt the framework ... the more complicated you make it, the more knobs you provide developers, the more problematic it's going to be.''} Then, this friction would cause the developers to bypass the secure path. \textit{``maintenance tasks and onboarding tasks that your developers are going to have to do, every each step in your instruction manual, is a friction point that a lazy developer or what I call an average human, you know, is going to say - Hey, I don't know if I want to do this, or this is too complicated. I don't want to learn it (P5).''} Therefore, it was suggested to ponder trade-offs between policy structure that handles various contexts and the complexity that increases the burden to implement the policy system. 

If the benefits from the new installation are lower than the investment, it would de-motivate an organization to adopt the system, and vice versa. To that end, the adoption of advanced indoor sensing technology such as UWB or mmWave radar may require a compelling business case considering deployment costs. Therefore, it was discussed to take into account the impacts of financial and laborious burden for the adoption (e.g., maintenance) and benefits an organization could achieve from deploying and integrating such framework into their existing infrastructure. 

\subsection{Management Challenge}
The implementation and deployment are followed by the management to make the system run in practice. 
The organization's legal and compliance elements to handle risks are crucial for managing the system as it entails risk of \snp. To that end, our expert participants shared suggestions on legal and risk management from various angles, which we will dive deeper in this section. 

\subsubsection{Organizational legal and risk management}
To reduce risk of any security and privacy issues from new technology, it is common that an organization arranges their policy that handles a plethora of cases, requiring granular policy control. 

Participants suggested methods to start approaching the policy design process. For example, while granular policy control may be necessary to handle various contexts of users, it is recommended to employ a top-down approach which prioritizes the high level structure of the policy. Given that there are various contexts of XR spatial sharing applications, it would be hard to consider and find the `ultimate solution' to address every potential cases. Our expert participants discussed the benefits to simplify the policy management strategy while addressing the risk. As an example to do so, P6 shared a suggestion to default the highest confidentiality level originally designed for the ``worst case'' scenario where the risk caused by the data breach is the highest among use cases. In doing so, the policy could be prepared to handle the risk even if the ``worst'' incident happens: \textit{``You're not gonna necessarily want to have different use cases, different permissions, because it's gonna be hard to manage within the deployment of the solution. So you're probably gonna wanna take the highest classification and say, okay, worst case scenario, we're gonna be processing and displaying high confidentiality information.''} 

In addition, many participants recommended leveraging existing policy control originally designed for onboarding other technology. P6 pinpointed that, if XR spatial sharing technology has a good amount of overlap with the technology previously procured and pre-cleared, taking advantage of the existing framework could bring significant benefits: \textit{``The first one would be working with the current risk pillars, which are responsible for the implementation and the drafting of control objectives within the firm to understand if any of them need to be uplifted to include these use cases so that you're within the framework.''} and \textit{``we're talking about SaaS services and past services, things that are already on the market and that can easily be commercialized. Looking at from a biometrics perspective, things like solutions that are aimed towards verifying the validity or identity of individuals for the onboarding process ... These are all solutions that are quite well known within the market and easily understandable and explainable to our audit partners and our regulators.''} P3 echoed: \textit{``I would think that the compliance controls that are required for something like this would be similar to what we would have to apply to Zoom. So rather than you having to rebuild the whole thing, I would actually go to the product owners of the Zoom product and say, all right, what are your guys' requirements?''} This way, it could reduce the amount of work by avoiding designing the policy control from the scratch. Additionally, as the existing policy control has been reviewed and confirmed by the firm, it would be also easier for legal teams to review and approve the policy control design.  

Lastly, P2 also mentioned that data management requirements may differ based on jurisdiction (e.g., country by country) and it is essential to adhere to laws such as General Data Protection Regulation (GDPR), California Consumer Privacy Act (CCPA), and child privacy protection rules accordingly when it comes to implementation and data policy: \textit{``different jurisdictions will have like different laws regarding like how long you could keep this data ...  Maybe some say that they have a GDPR ... They keep the footage for seven days and things of that nature. And that could potentially affect how you do training in improving your algorithms or improving red teaming. Maybe if you're doing this in one jurisdiction that says you cannot be archiving any of this data at all, then you can't do your training on that subset of meetings that you take in this region.''} P6 further added \textit{``incident reporting and privacy reporting is very country-specific. So depending on the individuals which data was either compromised or impacted by the incident, you're going to see that different obligations apply depending on the countries''} Additionally, it was emphasized that having continuous communication with regulators via audit reports, routine inspections, and clear documentation is critical to gain early approval of new initiatives and ensuring compliance: \textit{``you have an incident or you have a data leak that gets reported into your regulator and then they start looking into this and...you have to explain what happened ... And then you're looking at different regulators from different countries coming in and looking at the setup and seeing whether or not this is gonna fit within the regulatory framework that they have implemented in the look out ... There are ways to manage the proactive communication of net new initiatives (P6).''}

\subsubsection{Enterprise's Policy Control for Exceptions}
Participants also highlighted the cost of managing exception cases that may lead to extra hurdles for task performance (e.g., seeking exception approval during meeting may cause extra wait time and hurt client experience.) Enterprise prevents \snp breach of confidential information by performing policy control and regulating employees' device usage. While acknowledging policy control enforcement is essential to prevent the potential breach, participants expressed their concerns over excessive costs and cumbersome processes that could impede employees' productivity and client engagement. P2 argued \textit{``Like you want to perhaps show me information about like my [wealth management] or something like that. You can't show it to me until you get it approved four to five days later, that could potentially lead to frustration and ultimately like the client maybe not engaging with the firm ... I would hope that they don't have to go through speaker clearance every time they meet with a client.''}. In summary, it is required to find the harmony between the policy enforcement and its labor cost. 

\subsubsection{User–Device Pairing and Control}
Participants described risks arising from mismatches between designated users and enterprise-owned MR devices, including unauthorized use and impersonation. Participants noted that shared or accessible devices can enable inadvertent disclosure or deliberate leakage. As 
P2 stated: \textit{``First question is how do you authenticate the user who is using the headset? Because maybe you could assign the headset to an employee, but maybe it gets misused by a different employee. Like what kinds of checks do you have in place?''} 
P7 echoed: 
\textit{``the kind of the threats, which could be, hey, we give super user access to these engineers. [Person A] is working with, sense of client data, but [person B] can log in as [A] and see what [A] is seeing..''} 
In response to these concerns, participants (P2, P7) emphasized controls that bind device operation to the intended user and session (e.g., per-user authentication, session binding, auto-lock on headset removal, role-based authorization). Participants (P3, P5) also mentioned about alignment with existing enterprise data-handling practices (e.g., encryption in transmission, access control and retention policies in storage) consistent with other data-collecting systems.
% Besides digital security and privacy, an organization is situated to manage security and privacy in physical space if an organization has physical workspaces. 
% Since the device belongs to enterprise, it could be exposed to not just a designated employee but also other employees. This means that other employees could access the device and impersonate the device owner to perform careless activity, sharing sensitive information in the office space. As P1 mentioned: \textit{``You are trusting the user that he will act in good faith ... whether he wants to share or not, unless he has some malicious thing to do ... He may go and chat, right? We cannot control it.''} P6 echoed \textit{``It's like your malicious actor is your employee and they're trying to leak that information.''} Owing to that, participants placed emphasis on employing rigorous authentication measures to ensure only authorized entities can access and use the device. 
% In addition, participants suggested to make sure to employ methods to secure data transmission (e.g,. encryption) and data storage (e.g., access control, retention policy), which also has been applied for existing data-collecting device systems, as demonstrated in the previous quotes. 

\subsubsection{Third-Party Integration and Contractual Obligation}
As it is common that an organization uses third-party service for their system building, participants also discussed several suggestions for third-party service integration. First, it is suggested to offload some of the obligations to third-party vendors that the enterprise partners with. For example, P3 mentioned \textit{``So if we're using the third parties, then there's more contractual obligations, right, about what it is that they would need to implement, right? So all the controls that we would say you need to implement, that would be the obligation of the third party that you would need to ensure they're implemented...So for example, if we're requiring encryption of some of our data then that's an obligation of the third party.''} P6 echoed \textit{``depending on how much time you have to implement all of the controls ... we would also look at dependencies on third parties''} 

However, participants discussed items to beware of for the third-party service adoption. For example, P6 argued that the organization needs to verify whether current requirements are adequate for technology newly introduced to the firm, XR spatial sharing feature in this context, if risk identification and mitigation strategies designed for the new technology need to be created. Additionally, P3 pointed that the involvement of third-party will add contractual demands (e.g., data segregation between an organization and third-party, incident notification), increasing the complexity on the organization's legal landscape: \textit{``But it's everything from the third party's obligation to notify us when maybe there's some kind of a cybersecurity breach within their organization, or contractual obligations associated with data protection, data segregation, service level agreements ... So there's lots of different things that they would need to provide to us to get us comfortable that their products are secure and not introducing vulnerabilities into our environment. Yeah, there are obligations for them to maintain it, to adhere to our lifecycle management.''} As such, enterprise should be cautious when deciding when and how to leverage third-party obligations and policies, and how that interplays with existing systems within.

%% "different nuances depending on the type of, depending on the implementation, right? So lots of, the whole legal challenge becomes more complex when you introduce the third party"
\section{Discussion}

%% rough writing
% Many discussions are centered around how to mitigate any potential risk from different angles. 

The system integration would be related to many different stakeholders, which requires the balance between various factors affecting different stakeholders. Therefore, we reflect on the considerations for various factors and, in turn, how to balance each stakeholder's demand. 

% clear communication
%% users: need to communicate what they want to share in prior
%% companies: need to communicate what is restricted in prior and what is happening in real time. 

% negotation is possible but not always
%% users: need to understand from companies, what's prioritized
%% companies: help users have capabilities for their S&P practice

% strategic decision is needed for the deployment
%% financial---installation + maintenance, on-boarding process.
%% implementation complexity --- developer

\subsection{Clear and Timely Policy Communication between Users and Enterprise}
\label{subsection:timelycommunication}
Many experts put an emphasis on the need of clear and timely communications between users and enterprise. First, it is important to help users to perform `good' practice to reduce \snp risks at the right timing. ``Users are not the enemy \cite{adams1999users}.'' While the security system is well built, the system could go wrong even if the user makes one single \snp decision mistake. The Security \& Privacy Acceptance Framework (SPAF) identifies three key factors to help users' to follow \snp practices while improving knowledge, awareness, and ability \cite{das2022security}. Additionally, doing so in a timely manner will amplify users' good \snp practices. As enterprise has often adopted those three factors to promote \snp practices to employees, they can provide the similar resources for MR spatial sharing. For example, educational materials regarding \snp risks can be provided to users prior to usage (knowledge),  effective nudge on the spatial sharing user interface when entering restricted area can be enabled to enhance the user's awareness about their activity (awareness), and \snp features like mute video on demand can be easily designed for users to control to perform \snp practices (ability). In summary, the clear guidance will help employees motivated to perform \snp practices to mitigate the risk.

On the other hand, users need to clearly communicate their demands regarding \snp practices with enterprise. However, it is recommended to do so in a timely way if possible. As mentioned in \Cref{section:results}, it is advised that employees can request approval for which physical contents to share prior to the meetings including spatial sharing which allows enterprise to have enough time performing a thorough evaluation for the request. 

On the technical side, in addition to these communication strategies, it is valuable to ensure that policy documentation is readily accessible to users within the MR system itself. By embedding clear and concise policy information directly into the spatial sharing interface, employees can quickly reference relevant guidelines and requirements at the moment they need them. This approach supports users in making informed decisions, reinforces awareness of enterprise expectations, and reduces the likelihood of accidental policy violations. Providing in-system access to policy documentation complements educational materials and real-time nudges, further empowering users to follow best practices for \snp in their daily workflows.

Taken together, clear and timely communication is essential to meet the expectations from both users and enterprise ends to perform \snp practices. 
 
\subsection{Negotiation is Possible But Not Always}

A user may want to show a particular physical object or area which is not explicitly guided by enterprise. For such cases, enterprise can provide a channel for users to make requests to add contents to share during their meeting unless contents to share are strictly restricted. As mentioned in \Cref{subsection:timelycommunication}, it is suggested to perform the clearance of sharing in prior, helping the enterprise to advise employees how to properly take actions for spatial sharing.

On the other hand, from the enterprise point of view, the primary goal is to reduce risk of data breach. To that end, experts suggested prioritizing the prevention of any sensitive data in physical space from accidentally being exposed. One suggested way is employing the `allowlist' approach rather than `blocklist' where `allowlist' is to block all the contents by default and allow a certain contents to be share per request and approval, and `blocklist' works otherwise. This way, while potentially detrimental for usability, the default setup to prevent any sensitive information from being shared unless requested would help reduce the data breach risk.
Therefore, users need to have a clear understanding of the role of the allowlist approach. 

In addition to prior clearance and the allowlist approach, it is also beneficial to support user-initiated overrides for cases where non-sensitive content is misclassified as sensitive by the system. By enabling users to request an override, with appropriate logging and approval workflows, enterprises can reduce unnecessary usability barriers to sharing legitimate information while maintaining accountability. However, to ensure that \snp are not compromised, override permissions should be context-dependent and more restrictive in high-risk environments. For example, in areas designated as highly sensitive, the system may still limit or prohibit overrides altogether, whereas in lower-risk contexts, users may be granted more flexibility. This balanced approach helps maintain robust protection of sensitive data while allowing for practical and efficient collaboration.

\subsection{Deployment Needs Strategic Decision}

Benefits over investment need to be thoroughly studied, some experts mentioned. For example, installation of new technology to the existing infrastructure would need financial investment, which may not be trivial from the enterprise perspective. In addition, the maintenance of the technology is necessary for employees to use over time, accumulating the financial burden simultaneously. To that end, it would be critical to justify the financial burden increased by the deployment and management of the newly introduced technology and assess the business value.

During the interview, we observed the concern that application developers may feel de-motivated to build the system if it is hard to implement the system design. For example, there is technical difficulties to implement secure system by following the recommended ways to improve \snp, echoing the developers' challenges identified by Li at al. \cite{li2021developers}. Therefore, instead of forcing them to build whatever requests were made from different stakeholders, it would be critical to communicate and closely work with developers to understand their implementation challenges and help them to be motivated to build the critical systems for \snp.

We also recommend support developers and streamline the deployment process by providing developer essential toolkits for each module implementation, along with clear documentation, sample code, or template policy build blocks to facilitate smoother integration and encourage best practices. This may as well help the developers to better understand the \snp controls and make it easier for developers to implement local policies for end-user applications. These efforts are well aligned with, and build upon, prior findings that demonstrate the effectiveness of providing clear documentation and visualization tools to developers with less \snp expertise\cite{Rajaram2025}. Additionally, iterative onboarding, where advanced features and concepts are introduced gradually rather than all at once, can further minimize initial complexity and help developers build confidence as they work with the system.

There is always a possibility that a system malfunctions. Even during this incident, it would be important to minimize the usability issue caused by the malfunctions. To do so, the system structure needs to be strategically designed. One example could be to delegate camera filtering operation to a local device instead of the reliance on server performing the camera filtering. The camera filtering is one key feature to protect \snp of physical space information. However, running the feature on server may cause the latency of the view, in turn, causing frustrations. Then, they may find an `easy' way to perform spatial sharing, which could be more vulnerable to data breach. On the other, if the camera filtering runs on a local device, it could reduce the latency, enhancing the usability, and potentially preventing the users' frustration and stop them from finding the `easy' way. 

\subsection{Context-Aware Access Control and Device Security Are Foundational}

As experts highlighted, the ability to reliably determine a user’s physical location and context using different modalities such as RFID, UWB, mmWave, WiFi, or integration with enterprise badge and identification systems enables the system to dynamically enforce access policies as users transition between permitted and non-permitted areas. This real-time localization ensures that sensitive content and features are only accessible in appropriate contexts, reducing the risk of accidental exposure or unauthorized sharing. However, context-aware access control must be tightly coupled with robust device security measures to address the risks of unauthorized access and insider threats. As P1 and P6 mentioned about the potential for device impersonation and malicious insiders, it is crucial to implement rigorous authentication mechanisms such as multi-factor authentication (MFA), biometrics, and enterprise identity management integration. These controls help ensure that only authorized users can access MR devices and sensitive data, and that access permissions are continuously aligned with the user’s current context and location.

In addition, secure data transmission and storage are critical components of this security framework. As P3 discussed regarding authentication and data protection, all communications between MR devices, servers, and third-party services should be encrypted using industry-standard protocols. Data stored on devices or in the cloud should also be protected with robust encryption and granular access controls, ensuring that sensitive information remains secure throughout its lifecycle and is only accessible to users with the appropriate context and credentials.

By combining accurate context-aware localization with strong authentication and comprehensive data protection measures, enterprises can establish a resilient foundation for secure and privacy-preserving MR spatial sharing. This integrated approach not only mitigates the risk of unauthorized access and data breaches, but also supports compliance with enterprise policies and regulatory requirements, ensuring that security is maintained as users move through diverse physical and digital environments.

\subsection{Proactive Jurisdiction-specific Legal Compliance and Device Security Are Essential}

The deployment of MR spatial sharing systems in enterprise environments introduces complex legal and compliance challenges, especially for organizations operating across multiple jurisdictions. As P2 mentioned about varying data retention laws, it is essential to implement jurisdiction-aware data management that allows configurable data retention, access, and reporting policies tailored to the requirements of regulations such as GDPR, CCPA, and other local privacy frameworks. This adaptability ensures ongoing compliance and enables organizations to respond effectively to changes in legal standards. To further support compliance, organizations should incorporate automated compliance checks that regularly audit system configurations and data flows for adherence to both regulatory and internal policy requirements. As P6 noted about the country-specific nature of incident and privacy reporting, automated mechanisms can help generate jurisdiction-specific reports and maintain a robust audit trail, which is critical for regulatory reviews and investigations. Additionally, proactive regulator engagement is also recommended, as highlighted by P6 regarding the importance of continuous communication with regulators through audit reports, routine inspections, and clear documentation. Establishing transparent documentation and open communication channels can facilitate smoother approval processes for new MR technologies and foster trust with oversight authorities.

In terms of device security, several experts emphasized the risk of unauthorized access and insider threats. As P1 and P6 mentioned about the potential for device impersonation and malicious insiders, it is crucial to implement rigorous authentication measures such as multi-factor authentication (MFA), biometrics, and integration with enterprise identity management systems. These controls not only help ensure that only authorized users can access MR devices and sensitive data, but also play a critical role in meeting regulatory requirements for access control and auditability. Failure to adequately secure devices can result in unauthorized disclosures, which may trigger significant legal consequences, including regulatory investigations, fines, and reputational damage for the enterprise.

Furthermore, secure data transmission and storage must be prioritized as part of a comprehensive legal and compliance strategy. As P3 discussed regarding authentication and data protection, all communications between MR devices, servers, and third-party services should be encrypted using industry-standard protocols. Data stored on devices or in the cloud should also be protected with robust encryption and granular access controls, ensuring that sensitive information remains secure throughout its lifecycle and in accordance with applicable data protection laws. In the event of a security incident, such as a data breach or unauthorized access, organizations must be prepared to respond promptly and transparently, fulfilling incident reporting obligations as required by regulations in different jurisdictions. By integrating these technical safeguards with advanced compliance strategies and strong authentication, enterprises can establish a resilient foundation for secure and privacy-preserving MR spatial sharing that meets both operational and regulatory expectations, and is equipped to handle the legal ramifications of potential security incidents.

\subsection{UI/UX for Better Awareness and User Education / Training}

A recurring theme in the expert feedback was the value of familiar interface controls, such as mute or hide camera buttons, which empower users to quickly disable video feeds or prevent accidental sharing. These features not only give users more usability control but also serve as an additional layer of protection against system errors. However, the effectiveness of such controls depends on users being aware of what is currently being shared and understanding the underlying enterprise policies that govern these decisions. To address this, we recommend that future spatial privacy systems incorporate clear and context-sensitive interface notifications. These notifications should inform users about which objects or content are being shared or restricted, and provide concise explanations of the relevant enterprise policies. By making policy decisions transparent and easily accessible, notifications can help users make informed choices and reduce the likelihood of accidental exposure.

Beyond usability and security benefits, interface notifications also offer educational value. Prior work in usable \snp has shown that well-designed feedback and notifications can help users develop a better understanding of privacy risks and system functionality \cite{bravo-lillo2013your}. Real-time notifications that explain why certain content is blocked or flagged can prompt users to reflect on the sensitivity of information and the rationale behind organizational policies. Over time, these educational prompts can foster a culture of privacy awareness, encouraging employees to be more vigilant and proactive in protecting sensitive data. Additionally, such notifications can help demystify complex policy rules, making them more approachable and less likely lead to confusion or frustration \cite{schaub2015, cao2021, sunshine2009}. In this way, interface notifications serve as both a practical tool and a continuous learning opportunity, reinforcing best practices and supporting the organization’s broader \snp goals.

\section{Conclusion}
Multi-user MR spatial sharing in enterprise environments holds significant promise, but also introduces pressing \snp risks arising from complex interactions between users and organizational contexts. These challenges should be addressed proactively as MR technologies are integrated into enterprise workflows, rather than after ad hoc practices emerge. To that end, we are the first to systematically identify and analyze \snp considerations for MR spatial sharing in enterprise settings. Through formative studies and expert interviews, we developed a probe framework and surfaced key risks and requirements for secure, usable spatial sharing. By articulating actionable design recommendations, our work lays the groundwork for responsible and secure MR adoption in the enterprise, taking essential steps toward safeguarding future multi-user MR collaboration.

\bibliographystyle{plain}
{\small
\bibliography{ref}}

\section*{Disclaimer}
This paper was prepared for informational purposes by the Global Technology Applied Research center of JPMorganChase. This paper is not a product of the Research Department of JPMorganChase. or its affiliates. Neither JPMorganChase. nor any of its affiliates makes any explicit or implied representation or warranty and none of them accept any liability in connection with this paper, including, without limitation, with respect to the completeness, accuracy, or reliability of the information contained herein and the potential legal, compliance, tax, or accounting effects thereof. This document is not intended as investment research or investment advice, or as a recommendation, offer, or solicitation for the purchase or sale of any security, financial instrument, financial product or service, or to be used in any way for evaluating the merits of participating in any transaction.

\end{document}